\documentclass[12pt, draftclsnofoot, onecolumn]{IEEEtran}
\usepackage{epsfig,latexsym}
\usepackage{float}
\usepackage{indentfirst}
\usepackage{amsmath}
\usepackage{amssymb}
\usepackage{times}
\usepackage{subfigure}
\usepackage{psfrag}
\usepackage{hyperref}
\usepackage{cite}
\usepackage{lastpage}
\usepackage{fancyhdr}
\usepackage{color}
 \usepackage{amsthm}
\usepackage{bigints}
\begin{document}
\title{ {\huge  A Survey on Non-Orthogonal Multiple Access for 5G Networks: Research Challenges and \\Future Trends}}

\author{ Zhiguo Ding, \IEEEmembership{Senior Member, IEEE}, Xianfu Lei,   \IEEEmembership{Senior Member, IEEE}, George K. Karagiannidis, \IEEEmembership{Fellow, IEEE}, Robert Schober, \IEEEmembership{Fellow, IEEE}, Jihong Yuan, \IEEEmembership{Fellow, IEEE}, and Vijay Bhargava, \IEEEmembership{Life Fellow, IEEE} \thanks{
    Z. Ding is also with the School of
Computing and Communications, Lancaster
University, Lancaster, UK (email: \href{mailto:z.ding@lancaster.ac.uk}{z.ding@lancaster.ac.uk}).

X. Lei is with the Institute of Mobile
Communications, Southwest Jiaotong University, Chengdu, China (email: \href{mailto:xflei81@gmail.com}{xflei81@gmail.com}).

G. K. Karagiannidis is with the Department of Electrical and Computer
Engineering, Aristotle University of Thessaloniki, Thessaloniki, Greece (email: \href{mailto:geokarag@auth.gr}{geokarag@auth.gr}).

R. Schober is with the Institute for Digital Communications,
Friedrich-Alexander-University Erlangen-Nurnberg (FAU), Germany (email: \href{mailto:robert.schober@fau.de}{robert.schober@fau.de}).

 J. Yuan is with the
School of Electrical Engineering and Telecommunications, the University of
New South Wales, Australia (email: \href{mailto:jinhong@ee.unsw.edu.au}{jinhong@ee.unsw.edu.au}).

V. Bhargava is  the Department of Electrical and Computer
Engineering, University of British Columbia, Vancouver, Canada (email: \href{mailto:vijayb@ece.ubc.ca}{vijayb@ece.ubc.ca}).
}\vspace{-1em}} \maketitle
\begin{abstract}
Non-orthogonal multiple access (NOMA) is an essential enabling technology for the fifth generation (5G) wireless networks to meet the heterogeneous demands on low latency, high reliability, massive connectivity, improved fairness, and high throughput. The key idea behind  NOMA is to serve multiple users in the same resource block, such as a time slot, subcarrier, or spreading code. The NOMA principle  is a general framework, and several  recently proposed 5G multiple access schemes can be viewed as special cases. This survey provides an overview  of the latest NOMA research  and innovations as well as their applications. Thereby, the papers published in this special issue are put into the content of the existing literature.  Future research challenges regarding  NOMA in 5G and beyond are also discussed.
\end{abstract}\vspace{-1em}
\section{Introduction}
Non-orthogonal multiple access (NOMA) has become an important  principle for the design of radio  access techniques for the fifth generation (5G) wireless networks \cite{nomama, 7263349, ztenoma}. Although  several  5G multiple access techniques have been proposed by academia and industry, including power-domain NOMA \cite{6692652,NOMAPIMRC,Nomading},  sparse code multiple access (SCMA) \cite{6666156, 6966170}, pattern division multiple access (PDMA) \cite{7024798,7526461},  low density spreading (LDS) \cite{6314235}, and lattice partition multiple access (LPMA) \cite{lpma}, these      techniques are based on the same key concept, where  more than one user is   served in each orthogonal  resource block, e.g., a time slot, a frequency channel, a spreading code, or an orthogonal spatial degree of freedom.

Unlike NOMA, conventional orthogonal multiple access (OMA) techniques, such as time division multiple access (TDMA) and orthogonal frequency division multiple access (OFDMA), serve a single user in each orthogonal resource block. The spectral inefficiency of OMA can be   illustrated with  the following simple example.  Consider a scenario, where   one user with very poor channel conditions needs to be served for   fairness purposes, e.g., this user has high priority data or has not been served for a long time.  In this case, the use of OMA means that it is inevitable that one of the    scarce bandwidth resources is solely occupied by this  user, despite its   poor channel conditions. Obviously, this has a negative impact on the spectrum efficiency and  throughput of the overall system. In such a situation,  the use of NOMA ensures not only that the user with poor channel conditions is served but also that   users with  better channel conditions  can concurrently  utilize  the same bandwidth resources as  the weak user. As a result, if user fairness has to be    guaranteed, the system throughput of NOMA can be significantly larger than that of OMA \cite{Zhiguo_CRconoma}.  In addition to its spectral efficiency gain, academic and industrial research has  also demonstrated that NOMA can effectively support massive connectivity, which is important for ensuring  that the forthcoming 5G network can support the Internet of Things (IoT) functionalities \cite{docom, huawei, NGMN,teckuk}. 

Although the application of NOMA in cellular networks is relatively new, related concepts have been studied in information theory for a long time. For example,   key components  of NOMA, such as   superposition coding, successive interference cancellation (SIC), and the message passing algorithm (MPA), have already been invented more than two decades ago \cite{xiaobook,Verduebook}.   Nevertheless,  the   principle  of NOMA, i.e., removing orthogonality, has not been used in the previous generations of cellular networks.  In this content, we note that the philosophy behind NOMA is rarther different from that behind   code division multiple access (CDMA). In fact,  CDMA is primarily built upon  the idea that users  are separated   by exploiting the differences among  their   spreading  codes, whereas NOMA encourages multiple users to employ       exactly the  same code. As a consequence, for CDMA, the chip rate has to be much higher than the  supported information data rate, e.g., supporting a  data rate of 10 Gbps may require  a chip rate of a few hundred Gbps,  which is difficult to  realize  with practical hardware.

Conventionally, NOMA can be integrated in existing and future wireless systems because of its compatibility  with  other communication technologies. For example, NOMA has been shown to be compatible with conventional OMA, such as TDMA and OFDMA \cite{Rappaport}. Because of this, NOMA has also been  proposed for inclusion in  the  3rd generation partnership project (3GPP) long-term evolution advanced (LTE-A) standard \cite{3gpp1}, where NOMA is referred to as multi-user superposition transmission (MUST). Particularly, without requiring any changes to the LTE resource blocks (i.e.,  OFDMA subcarriers),  the use of the NOMA principle  ensures that two users are simultaneously served on the same OFDMA subcarrier.   Furthermore,  NOMA has been recently included in the forthcoming  digital TV standard (ATSC 3.0), where it is referred to as layered division multiplexing (LDM) \cite{7378924}.  Particularly,  the spectral efficiency of TV broadcasting is improved by  using  the NOMA principle and superimposing multiple data streams. The above examples clearly demonstrate the large potential of NOMA, not only for 5G networks, but also for  other upcoming and existing wireless systems.

The goal  of this survey is to provide a comprehensive overview   of the latest NOMA research results and innovations, including  the papers published in this JSAC special issue. In particular, the design of single-carrier and multi-carrier NOMA is discussed  in Sections \ref{section single carrier} and  \ref{mc noma}, respectively, where power-domain NOMA, SCMA, LDS, and PDMA are used as examples. Sections \ref{section mimonoma} and \ref{section cooperative noma} focus on multiple-input multiple-output (MIMO) NOMA and cooperative NOMA, respectively. In Section \ref{section mmwave noma}, the combination of NOMA with millimeter-wave (mmWave) communications is studied. Some important implementation issues of NOMA are discussed in Section \ref{section implementation},  and  concluding remarks are provided in Section \ref{section conclusion}.

\section{Single-Carrier NOMA}\label{section single carrier}
 When  the NOMA principle is applied to a single orthogonal resource block, i.e.,   a single carrier, a spectrally efficient way  to realize multiple access is to utilize the power domain. This leads to so-called {\it power-domain NOMA} which will be discussed first in this section. Subsequently,  a variation of power-domain NOMA, which is referred to as {\it cognitive radio inspired NOMA} (CR-NOMA) and can strictly meet the users' diverse quality of service (QoS) requirements, is  described.
\subsection{Power-domain NOMA}\label{subsection 2:1}
\begin{figure}[thb]
\begin{center}
\includegraphics[width=8cm]{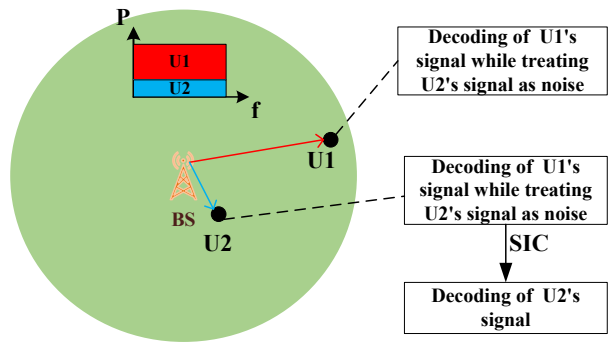}
\end{center}\vspace{-1em}
\caption{An example of a NOMA downlink scheme with  two users, denoted by U1 and U2, and  one subcarrier.  }
        \label{Fig:plt_CR7}
\end{figure}

Power-domain NOMA can serve multiple users in the same time slot, OFDMA subcarrier,  or spreading code, and    multiple access is realized by allocating different power levels to different users \cite{NOMAPIMRC,Nomading,7527668,Krikidisnoma,6708131}. An example for this version of NOMA is MUST, which is a two-user downlink   power-domain NOMA scheme, where a base station (BS) serves simultaneously two single-antenna users at the same   OFDMA subcarrier. Denote the two users' channels by $h_i$, $i\in\{1,2\}$, and assume $|h_1|\leq |h_2|$. The BS   superimposes the users' messages by assigning corresponding  power coefficients, denoted by $\alpha_i$, $i\in\{1,2\}$. As shown in Fig. \ref{Fig:plt_CR7}, the key idea of power-domain NOMA is to allocate more power to the user with poorer  channel conditions, i.e., $\alpha_1\geq \alpha_2$ and $\alpha_1^2+\alpha_2^2=1$ if $|h_1|\leq |h_2|$. User 1 decodes its own message directly by treating user $2$'s message as noise, which results in an achievable rate of $\log_2\left(1+\frac{|h_1|^2\alpha_1^2}{|h_1|^2\alpha_2^2+\frac{1}{\rho}}\right)$ bits/s/Hz, where $\rho$ denotes the transmit signal-to-noise ratio (SNR). On the other hand, user $2$ performs  SIC, i.e., it first decodes user $1$'s message and then removes this message from its observation before decoding its own message. This   strategy results in an achievable rate of $\log_2\left(1+\rho|h_2|^2\alpha_2^2\right)$ since SIC is always possible as $\log_2\left(1+\frac{|h_2|^2\alpha_1^2}{|h_2|^2\alpha_2^2+\frac{1}{\rho}}\right)\geq \log_2\left(1+\frac{|h_1|^2\alpha_1^2}{|h_1|^2\alpha_2^2+\frac{1}{\rho}}\right)$.

The spectral efficiency  gain of NOMA can be illustrated in on the following examples.

 {\it Example 1:} Consider a high-SNR scenario, i.e., $\rho\rightarrow \infty$. For   illustration purpose, assume $\rho |h_1|^2\rightarrow 0$, i.e., user $1$'s channel experiences a deep fade. The sum rate achieved by NOMA can be approximated as follows:
    \begin{align}\label{sumrateX1}
    &\log_2\left(1+\frac{|h_1|^2\alpha_1^2}{|h_1|^2\alpha_2^2+\frac{1}{\rho}}\right)+\log_2\left(1+\rho|h_2|^2\alpha_2^2\right)\\\nonumber
    \approx&\log_2\left(1+\frac{ \alpha_1^2}{ \alpha_2^2}\right)+\log_2\left(\rho|h_2|^2\alpha_2^2\right)=\log_2(\rho |h_2|^2).
    \end{align}
On the other hand, the sum rate of OMA can be expressed as follows:
    \begin{align}\label{sumrateX2}
    &\frac{1}{2}\log_2\left(1+\rho|h_1|^2\right)+\frac{1}{2}\log_2\left(1+\rho|h_2|^2\right)\\\nonumber
    \approx&\frac{1}{2}\log_2\left(\rho|h_2|^2\right).
    \end{align}
    The performance gain of NOMA over OMA is obvious from \eqref{sumrateX1} and \eqref{sumrateX2}.

     {\it Example 2:} Assume that user $1$ is an IoT device requiring only a low data rate, and user $2$ is a user demanding a high data rate. When OFDMA is used, which is a typical example of OMA,    each user is allocated   one subcarrier. In this example,  the spectral efficiency of OMA is poor since the IoT device is served with  more bandwidth than what it actually needs, while  the broadband user is not assigned  enough bandwidth. On the other hand, the use of NOMA encourages  spectrum sharing, i.e., the broadband user can also have access to the subcarrier occupied by the IoT device. As a result, the use of NOMA  efficiently  supports massive connectivity and  meets the users' diverse QoS requirements \cite{Zhiguo_iot}.

         One may argue that the use of optimal resource allocation for OMA can overcome the above described disadvantage. However, as shown in \cite{zhiyongrui}, if both NOMA and OMA employ optimal resource allocation, NOMA   still yields significant performance gains. In addition, it is worth pointing out that adaptive resource allocation for OMA introduces  dynamic changes to the properties of the orthogonal resource blocks, and may require using   time slots with very short durations, which might not be realistic in practice.

\subsection{CR-NOMA}
Conventional power-domain NOMA  allocates more power to the user with poor channel conditions, which  ensures   user fairness; however,   conventional power-domain NOMA   cannot strictly guarantee the users' QoS targets. Cognitive-radio (CR) NOMA is an important variation of power-domain NOMA, which  strictly ensures that some or all of the users' QoS requirements are met. The key idea behind  CR-NOMA is to treat NOMA as a special case of cognitive ratio, where the power allocation policy is designed such that  the users' predefined QoS requirements are met.

  The scheme proposed in \cite{Zhiguo_CRconoma} is an example for CR-NOMA, where a BS   serves two downlink users by using the NOMA principle. Particularly, the user with the  poorer  channel conditions is viewed as the  primary user of a cognitive radio network, i.e., this user has a strict data rate requirement, which needs to be fulfilled. This data rate requirement imposes a constraint on the power allocation  policy   as follows:
\begin{align}
\log_2\left(1+\frac{|h_1|^2\alpha_1^2}{|h_1|^2\alpha_2^2+\frac{1}{\rho}}\right)\geq R_1,
\end{align}
where $R_1$ denotes user $1$'s target data rate. Considering   the above constraint, the cognitive radio inspired power allocation policy can be expressed as follows:
\begin{align}
\alpha_1^2=\min \left\{1,  \frac{\left(|h_1|^2+\frac{1}{\rho}\right)(2^{R_1}-1)}{|h_1|^2 2^{R_1}}\right\},
\end{align}
which means that all the power is allocated to user $1$ for a large $R_1$.

The rationale behind the above CR-NOMA power allocation policy is that user $1$ will be always served with   sufficient power  to satisfy its QoS requirements. If there is any power left afterwards, user $2$ will be served by using the remaining power. The benefits of CR-NOMA can be explained based on Example $2$ described in the previous subsection. Particularly, the use of OMA implies  that one subcarrier is solely occupied  by the low-rate IoT device. By using CR-NOMA, not only the IoT device can be served  with its targeted QoS requirement, but also one additional user can be admitted to this subcarrier, which increases the overall system throughput.

The outage and rate performance of CR-NOMA is analyzed in \cite{Zhiguo_CRconoma} for the case of two scheduled users, while the energy efficiency of CR-NOMA is studied in \cite{7794658}, where the concept of CR-NOMA is extended to  systems  with multi-antenna nodes. We note that it is not necessary to always treat the user with poor channel conditions as the primary user. In \cite{7542118}, a more general cognitive radio inspired power allocation policy is proposed for downlink and uplink NOMA scenarios,    in order to meet all  users'   QoS requirements in a more flexible manner.

\section{Multi-Carrier NOMA}\label{mc noma}
Given the technical maturity of OFDMA, this type of OMA will very likely be incorporated into 5G networks, and therefore how multiple OFDMA subcarriers can be efficiently combined with NOMA has received a lot of attention.  In this section, first the general principle of multi-carrier NOMA is introduced, and then several  existing forms of  multi-carrier NOMA   proposed for 5G networks are  described.
\subsection{Multi-carrier NOMA: A special  case of hybrid NOMA}
Multi-carrier NOMA can be viewed as a variation  of hybrid NOMA, where the users in a network are divided into multiple groups. Particularly,  the users in each group are served in the same orthogonal resource block following the NOMA principle, and different groups are allocated to different orthogonal resource blocks. The motivation for employing hybrid  NOMA is to reduce the system complexity. For example, assigning  all the users in the network  to a single group for the implementation of NOMA in one orthogonal resource block is problematic, since the user having the best channel conditions will have to decode all the other users' messages before decoding its own message, which results in high complexity and high decoding delay. Hybrid NOMA is an effective approach to strike a balanced tradeoff between system performance and complexity. Let's consider  multi-carrier NOMA as an example. The users in the cell are divided into multiple groups which are not necessarily mutually exclusive. The users within one group are allocated to  the same subcarrier, and  intra-group interference is mitigated  by using the NOMA principle. Different groups of users are allocated to different subcarriers, which   effectively avoids inter-group interference. As a result, overloading the system, which is necessary  in order to support more users than the number of available subcarriers and is required to enable massive connectivity,  can be still realized by this hybrid NOMA scheme. It is noted  that, with hybrid NOMA, overloading is realized at reduced  complexity since  the number of users at each subcarrier is limited.  Compared to other forms of hybrid NOMA, multi-carrier NOMA is particularly attractive, mainly due to the fact that OFDMA is likely to be employed in the forthcoming 5G network.

A key step to implement hybrid NOMA is to understand the impact of user grouping on the system performance. User pairing, i.e., assigning  two users to a single orthogonal resource block, is studied in \cite{Zhiguo_CRconoma}. In particular, this paper  demonstrates that grouping those two users from a set of users, whose channel conditions are most different,   yields  the highest   performance gain over  OMA. It is noted that the use of NOMA has quite different effects on the two users. For example, the user with stronger channel conditions prefers NOMA, since it is very likely that this  user achieves  a higher  individual data rate with NOMA   compared to   OMA. However, the  NOMA rate of the user with poorer channel conditions can be smaller than the rate with  OMA, where this performance loss can be mitigated by the CR inspired approach discussed in the previous section. Nevertheless,  the sum rate achieved with NOMA is typically much larger than that with OMA, particularly if  two users with quite different  channel conditions are paired.

Practical design strategies  for multi-carrier NOMA are also important to ensure that the performance gains predicted by the aforementioned  theoretical studies regarding  user grouping are realized in practice. Developing corresponding resource allocation algorithms is  challenging as shown in \cite{7812683, jsacnoma24, Robertnoma}, since   the difficult problems of user grouping,  subcarrier allocation, and power allocation are coupled. Monotonic optimization is applied to solve  the resulting  non-convex optimization  problem in \cite{7812683}, where an optimal solution for joint subcarrier allocation, user grouping, and power allocation is obtained.  This optimal solution  is important since it provides an algorithmic upper bound for the performance of multi-carrier NOMA. A low-complexity suboptimal solution based on successive convex optimization is also proposed in \cite{7812683}, and the corresponding resource allocation algorithm     achieves  a performance gain close to that of the optimal one.

\subsection{LDS, SCMA, and PDMA}
Since  multi-carrier NOMA achieves  a favourable  tradeoff between system performance and complexity, various practical forms of multi-carrier NOMA have been proposed for the 5G standard. Both LDS and SCMA are based on the   idea that one user's information is spread over multiple subcarriers \cite{6666156, 6966170,6314235,7561515x}. However,  the number of  subcarriers assigned to each user is smaller than  the total number of subcarriers, and this low spreading (sparse) feature  ensures that the number of users utilizing the same subcarrier is not  too large, such that the system complexity remains  manageable. This is illustrated in the following example.

\begin{figure}[thb]
\begin{center}
\includegraphics[width=9cm]{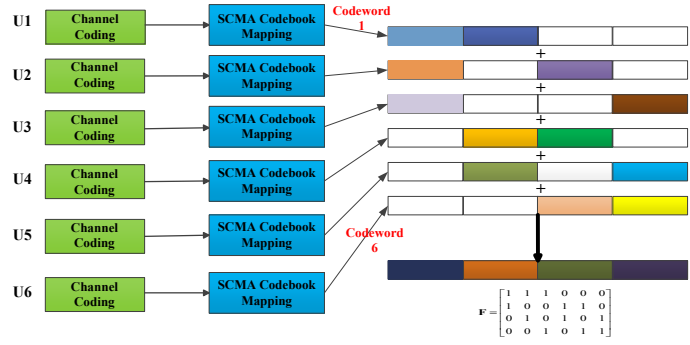}
\end{center}\vspace{-1em}
\caption{An example of an  SCMA system  with  six users and  four subcarriers.  }
        \label{Fig:plt_CR6}
\end{figure}

  {\it Example 3}: Consider an SCMA system  with $6$ users and $4$ subcarriers, as shown in Fig. \ref{Fig:plt_CR6}.  The key step to implement SCMA is to design the factor graph matrix, which specifies  which user's encoded messages are allocated to which subcarriers \cite{7582475}. A typical factor graph matrix for SCMA with  $6$ users  and $4$ subcarriers   is   the following
\begin{align}\label{F}
\mathbf{F}=\begin{bmatrix}1 & 1 &1 & 0 &0 &0 \\1 & 0 &0 & 1 &1 &0 \\0 & 1 &0 & 1 &0 &1 \\0 & 0 &1 & 0 &1 &1\end{bmatrix},
\end{align}
where $[\mathbf{F}]_{i,j}=1$ means that the $j$-th user can use the $i$-th subcarrier, and $[\mathbf{F}]_{i,j}=0$ means that this user cannot use the subcarrier. The sparse feature of SCMA is reflected in the fact that there are only two non-zero entries in each column of $\mathbf{F}$, i.e., each user employs   only two subcarriers. Since one user can use multiple subcarriers, SCMA  employs  multi-dimensional coding in order to ensure that the user's information is effectively spread over the subcarriers.  Because  one user's messages at different subcarriers are jointly encoded, SCMA requires joint decoding at the receiver, where the MPA is used to ensure low complexity \cite{7208827,7567516, 7467454}. This is an important  feature of SCMA, which distinguishes  it from power-domain NOMA, as joint decoding instead of SIC is employed.

PDMA can be viewed as another type of multi-carrier NOMA, but the low density spreading (sparse) feature of LDS and SCMA is no longer  strictly present, i.e.,  the number of  subcarriers occupied by one user is not necessarily much smaller than the total number of subcarriers. Similar to the factor graph matrix for SCMA, the performance of PDMA is largely determined by the design of the subcarrier allocation matrix, referred to as the   PDMA pattern matrix. This is explained   in the following example.

  {\it Example 4}:
  Consider a typical example for the implementation of PDMA, as proposed   in \cite{7024798,7526461}, where there are five users and three subcarriers. For this particular example, the PDMA pattern  matrix can be chosen as
\begin{align}\label{P}
\mathbf{Q}=\begin{bmatrix}1 & 1  & 0 &0 &1 \\1 & 1 &1 & 0  &0 \\1 & 0 &1 & 1 &0 \end{bmatrix},
\end{align}
where the entries of this matrix indicate how the subcarriers are allocated to the users, similar to $\mathbf{F}$ in \eqref{F}. However, different from LDS and SCMA, some users might be able to use all the subcarriers. For the example  in \eqref{P}, user $1$ is able to transmit or receive on all  subcarriers.

\section{MIMO-NOMA}\label{section mimonoma}
The application of MIMO to NOMA is   important, since the spatial degrees of freedom enabled by MIMO are crucial for meeting the performance requirements of  5G networks. This section presents   different approaches for the design of MIMO-NOMA.
\subsection{General principles of MIMO-NOMA}
\begin{figure}[thb]
\begin{center}
\includegraphics[width=6cm]{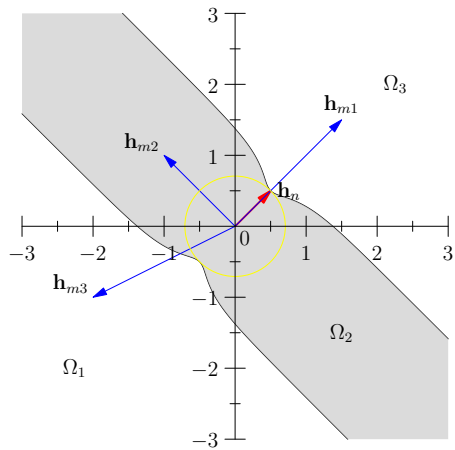}
\end{center}
\caption{An illustration of quasi-degradation when a BS  equipped with two antennas communicates with two single-antenna users. }
        \label{Fig:plt_CR4}
\end{figure}

  Compared to the design of single-input single-output (SISO) NOMA, the design of MIMO-NOMA is more challenging, mainly for  the following two reasons.  Firstly, it is not clear whether the use of MIMO-NOMA can achieve   the optimal system performance, although it is clear that MIMO-NOMA outperforms MIMO-OMA \cite{7465730}. Considering   downlink NOMA as an example, for the SISO case it is clear that the use of NOMA can realize a part of the capacity region of the broadcast channel. The probability for NOMA to achieve  larger individual and larger sum rates than OMA is rigorously obtained   in \cite{7272042}. However, the performance evaluation of MIMO-NOMA is more challenging. For a two-user downlink scenario with single-antenna users and a multiple-antenna BS,  \cite{7456242} and \cite{7555306} provide an information theoretic comparison between  NOMA, zero forcing, which can be viewed as a spatial OMA scheme,  and dirty paper coding (DPC), which achieves the capacity region of the broadcast channel with prohibitively high complexity \cite{Cover1991}. More importantly,   \cite{7456242} develops    the concept of quasi-degradation, which is a useful criterion for the evaluation  of MIMO-NOMA. This quasi-degradation concept is illustrated in Fig. \ref{Fig:plt_CR4}, where the BS has two antennas and there are two single-antenna users. Here, $\mathbf{h}_n$ denotes the $2\times 1$ channel vector of user $n$ which is paired with user $m$. Three possible realizations  of user $m$'s channel vector $\mathbf{h}_m$ are shown in the figure.  An extreme example for quasi-degradation is that one user's channel vector is a scaled version of the other user's channel vector, i.e., $\mathbf{h}_n$ and $\mathbf{h}_{m1}$ as shown in Fig. \ref{Fig:plt_CR4}. An extreme example for non-quasi-degradation is that the users' channel vectors are mutually orthogonal,  i.e., $\mathbf{h}_n$ and $\mathbf{h}_{m2}$ as shown in Fig. \ref{Fig:plt_CR4}. In general, if user $m$'s channel vector falls into the non-shaded area of $\Omega_1\cup\Omega_3$, the two users' channels are quasi-degraded. If the users' channels are quasi-degraded, it is shown in \cite{7456242} that the use of MIMO-NOMA can yield the same performance as DPC, i.e., the use of NOMA achieves the optimal performance in the MIMO context. We note that the extension of quasi-degradation to   general   multi-user MIMO   is still an open problem.

  Secondly, user ordering in MIMO-NOMA scenarios is a  difficult task.  In the SISO case, the users' channels are scalers, so it is straightforward to order the users according to their channel conditions. However, when nodes are equipped with multiple antennas, the users' channels are in form of vectors or matrices, which means that ordering users according to their channel conditions as in the SISO case becomes difficult. The use of random beamforming for NOMA was considered in \cite{6692307}, which   cleverly avoids the user ordering issue by asking  the BS to order the users according to their channel quality feedback. The use of  the large scale path loss for user ordering has also been proved to be an effective technique \cite{7095538, Zhiguo_faiconoma, jsacnoma2, jsacnoma21}. The scheme proposed in \cite{Zhiguo_faiconoma} is a representative  example for this approach, where users with larger distances from the BS are treated as {\it weak users} and their messages need to be decoded first at all the receivers. This user ordering strategy   results in a constraint for the system throughput maximization problem. Similarly, \cite{7582424} also employs    the users' path loss as indication for  their channel conditions.

\subsection{Decomposing MIMO-NOMA to SISO-NOMA}
Motivated by the user ordering difficulty, another effective approach to combining MIMO and NOMA  is to exploit spatial degrees of freedom and decompose MIMO-NOMA into multiple separate SISO-NOMA subchannels, which can significantly reduce the complexity of the system design \cite{Zhiguo_mimoconoma, jsacnoma16}.  Take a  downlink MIMO-NOMA scenario with $2M$ users as an example. The $2M$ users are randomly grouped into $M$ groups, where the two users in group $i$ are denoted by user $i$ and user $i'$, respectively. Assume that the BS is equipped with $M$ antennas, while  the users have $N$ antennas. User $i$  observes the following
\begin{align}\label{mimo1}
\mathbf{y}_i = \mathbf{H}_i \mathbf{P}\mathbf{x} +\mathbf{n}_i,
\end{align}
where $\mathbf{H}_i$ denotes the $N\times M$ channel matrix, $\mathbf{P}$ is a precoding matrix, $\mathbf{x}=\begin{bmatrix}x_1 &\cdots, &x_M \end{bmatrix}^T$, and each $x_i$ is a NOMA mixture containing the two symbols  for the two users in the same pair.

In \cite{Zhiguo_mimoconoma},   the case where  the BS does not know the users' channel matrices is considered, which means that a constant precoding matrix has to be used, e.g., the identity matrix can be used, $\mathbf{P}=\mathbf{I}_M$. As a result, each user applies zero forcing detection. Denote the detection vector by $\mathbf{v}_i$ and $\mathbf{v}_i^H\mathbf{p}_j=0$, for $i\neq j$, where $\mathbf{p}_j$ is the $j$-th column of $\mathbf{P}$.   Therefore, the resulting system model for the user can be written as follows:
\begin{align}\label{siso}
\mathbf{v}_i^H\mathbf{y}_i = \mathbf{v}_i^H\mathbf{h}_i  x_i +\mathbf{v}_i^H\mathbf{n}_i,
\end{align}
which corresponds to  a SISO-NOMA system model. Then, various SISO-NOMA designs can be applied in a straightforward manner. Note that $N\geq M$ is required for the scheme proposed in \cite{Zhiguo_mimoconoma} since the system relies on the receivers to suppress  inter-group interference.

When the users' channel matrices are known at the BS, \cite{Zhiguo_MIMONOMA2} shows that a better performance can be obtained. To avoid inter-group interference,   each precoding vector needs to satisfy the following constraint
\begin{align}\label{px}
\begin{bmatrix}\mathbf{v}_2^H \mathbf{H}_2
 &\mathbf{v}_{2'}^H \mathbf{H}_{2'} &\cdots & \mathbf{v}_{M'}^H \mathbf{H}_{M'}\end{bmatrix}_{2(M-1)\times M}\mathbf{p}_1=\mathbf{0},
\end{align}
where $\mathbf{p}_1$ is used as an example. Note that \eqref{px} is a set of   $2(M-1)$ linear equations with $M$ unknown variables ($\mathbf{p}_1$), which means that a solution does not exist.  Hence, the concept of signal alignment has been used in \cite{Zhiguo_MIMONOMA2}, as this facilitates zero forcing precoder design. Particularly, $\mathbf{v}_i$ and $\mathbf{v}_{i'}$ are chosen to ensure $\mathbf{v}_i^H \mathbf{H}_i=
 \mathbf{v}_{i'}^H \mathbf{H}_{i'}$. As a result, \eqref{px} can be simplified as follows:
 \begin{align}\label{px2}
\begin{bmatrix}\mathbf{v}_2^H \mathbf{H}_2  &\cdots & \mathbf{v}_{M}^H \mathbf{H}_{M}\end{bmatrix}_{(M-1)\times M}\mathbf{p}_1=\mathbf{0},
\end{align}
for  which a solution for $\mathbf{p}_i$ does exist. With such choices of $\mathbf{p}_i$ and $\mathbf{v}_i$, a  SISO-NOMA model similar to that  in \eqref{siso} is obtained, where the condition $N\geq M$ required in \cite{Zhiguo_MIMONOMA2} can be relaxed  to $N>\frac{M}{2}$. Note that the concept in \cite{Zhiguo_mimoconoma} and \cite{Zhiguo_MIMONOMA2} can be extended  to massive MIMO scenarios as shown in \cite{Zhiguo_massive}, by exploiting the spatial correlation among the users' channel matrices.

Other types of channel decomposition methods can be also applied to MIMO-NOMA, as shown in  \cite{7383326} and \cite{7504070}. For example, in \cite{7504070},  generalized singular-value decomposition   (GSVD) is applied in a NOMA network with two users. In order to simplify the illustration, assume  that all  nodes, including the BS and the two users, have $M$ antennas. Using GSVD, the two users' channel matrices can be decomposed as follows:
\begin{align}
\mathbf{H}_1 = \mathbf{U}_1 \mathbf{\Lambda}_1 \mathbf{Q}, \quad\quad \mathbf{H}_2 = \mathbf{U}_2 \mathbf{\Lambda}_2 \mathbf{Q},
\end{align}
where $\mathbf{U}_i$ is a unitary matrix, $\mathbf{\Lambda}_i$ is a diagonal matrix, and $\mathbf{Q}$ is an invertable  matrix. By using $\mathbf{Q}^{-1}$ as the precoding matrix, one can simultaneously diagonalize the two users' channel matrices. More importantly, the elements on the main diagonal of $\mathbf{\Lambda}_1$ are ascending and the elements on the main diagonal  of $\mathbf{\Lambda}_2$ are descending. Therefore, by using GSVD,   effective channel gains of different strength are paired together on each separated SISO channel,  which is ideal for the application of NOMA. 

It is worth pointing out that this approach of using spatial degrees of freedom can lead to a new form of NOMA, termed Angle Division Multiple Access (ADMA) as shown in \cite{jsacnoma6}. In addition, note that the use of antenna selection is another effective way to convert MIMO-NOMA to SISO-NOMA, where the maximum  diversity gain can be maintained at the price of a reduced multiplexing gain \cite{Sanayei04,anatenaselection, 7504208,jsacnoma27}. In general, these decomposition based MIMO-NOMA schemes not only reduce the complexity of the system design, but   are also general and applicable to both uplink and downlink transmission.

\subsection{When users have similar channel conditions}
The difference between the users' channel conditions plays an important role for the design of NOMA transmission, which  can be illustrated by using SISO-NOMA as an example. If two users have the same channels, i.e., $h_1=h_2$, the sum rate of NOMA in \eqref{sumrateX1} can be rewritten as follows:
   \begin{align}
    &\log_2\left(1+\frac{|h_1|^2\alpha_1^2}{|h_1|^2\alpha_2^2+\frac{1}{\rho}}\right)+\log_2\left(1+\rho|h_2|^2\alpha_2^2\right)
    =\log_2\left(1+|h_1|^2\right),
    \end{align}
    which is exactly identical to the sum rate of OMA.   Actually  many MIMO-NOMA designs rely on the assumption  that users have different path losses, and they cannot work properly when users have similar channel conditions. By considering \cite{Zhiguo_faiconoma} as an example, if the users have the same path loss, the feasible region of the NOMA rate optimization problem formulated in this paper may become empty.

For generalization of the NOMA principle, researchers have proposed  new forms of NOMA transmission. Both \cite{Zhiguo_iot} and \cite{Zhiguo_pls} order users not according to their channel gains but according to their  QoS requirements. Consider {\it Example 2} described in Section \ref{section single carrier} as a representative scenario. Assume that user $1$ has a low data rate requirement, such as an IoT device, and user $2$ is a broadband user. The use of OMA results in the situation that the IoT device is given more bandwidth than it needs, whereas the broadband user does not have enough bandwidth.  The use of NOMA can effectively avoid this problem. Take the scheme proposed in \cite{Zhiguo_iot} as an example. To simplify the illustration, we degrade the MIMO-NOMA scheme proposed in \cite{Zhiguo_iot} into  the simpler case with single-antenna users. The beamforming vector is matched to user $2$'s channel, which means that the two users' effective channel gains are $\frac{|\mathbf{h}_2^H\mathbf{h}_1|^2}{|\mathbf{h}_2|^2}$ and $|\mathbf{h}_2|^2$, respectively. There are two reasons for such a beamforming design. One is to ensure that user $2$'s effective channel gain is improved since this user needs to be served with a larger data rate, and the other is that this beamforming effectively  enlarges the difference between the two users' channel conditions, which is beneficial for the application of NOMA.  It is worth pointing out that although user $1$'s effective channel may not be strong, effective power allocation policies can also be used  to ensure that this user's QoS requirements are strictly met.

\section{Cooperative NOMA} \label{section cooperative noma}
In this section, the principle of cooperative NOMA is introduced and  two types of cooperation are considered.

\subsection{Cooperation among NOMA users}
\begin{figure}[thb]
\begin{center}
\includegraphics[width=10cm]{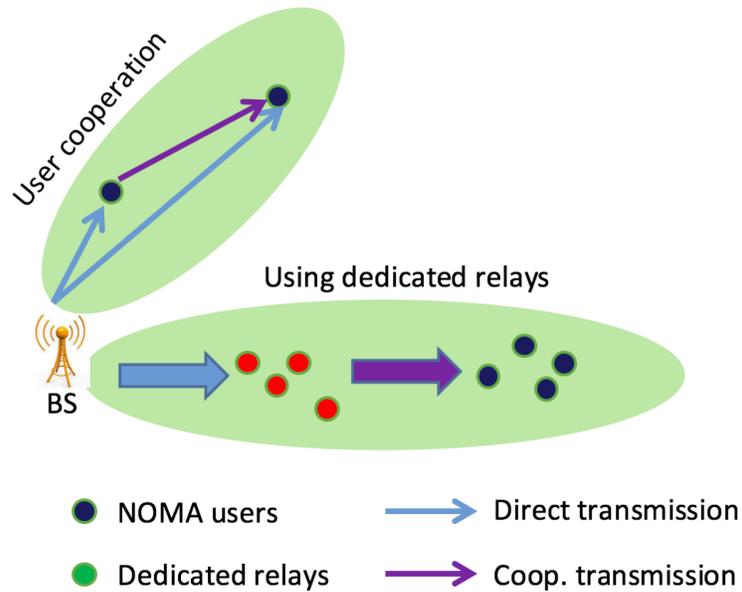}
\end{center}
\caption{An illustration of the two forms of cooperative NOMA.    }
        \label{Fig:plt_CR1}
\end{figure}

The first type of cooperative NOMA  considers the cooperation among the NOMA users, where one NOMA user acts as a relay for the other one, as shown in Fig. \ref{Fig:plt_CR1}. This type of cooperative NOMA is motivated by the following reasons:
\begin{itemize}
\item There is redundant information in NOMA systems, which can be employed for   cooperative transmission. The two-user downlink case is such  an example. The strong user needs to decode the weak user's information, before decoding its own signal. This means that the strong user can act as a regular relay, which assists  the weak user.

\item There is a need for carrying out cooperative transmission in NOMA systems. Take again {\it Example 1} described in Section \ref{subsection 2:1} as an example. User $1$, i.e., the weak user, has a rate of $\log_2\left(1+\frac{|h_1|^2\alpha_1^2}{|h_1|^2\alpha_2^2+\frac{1}{\rho}}\right)$, which is negatively affected by   the co-channel interference from the strong user. Cooperative transmission can improve the weak user's data  rate.
\end{itemize}

In \cite{Zhiguo_conoma}, a cooperative NOMA protocol    relying  on   cooperation among the NOMA users has been proposed. Particularly, consider a downlink transmission  example with two users, where cooperative NOMA transmission is performed in two phases. During the first phase, the BS broadcasts the superimposed mixture of the users' signals. In the second phase, the strong user acts as a relay and forwards the weak user's message to the weak user, by using short-range communications, such as Bluetooth or WiFi. Furthermore, it is shown in \cite{Zhiguo_conoma} that, even if short-range communications are not used, the performance of cooperative NOMA is still  superior to cooperative OMA. The reason for this is that cooperative NOMA requires two time slots only, if short range communications is not used. In contrast, cooperative OMA needs three time slots, where the BS uses  two time slots to deliver the two messages to the two users, respectively, and one additional time slot is needed for the strong user to assist  the weak user. The spectral efficiency of cooperative NOMA can be further improved by employing full duplexing relaying, as shown in \cite{7542601, 7572025, jsacnoma15}. In these works, by using full duplexing, the strong user receives the signals from the BS and carries out relay transmission simultaneously. This avoids the disadvantage of half-duplexing relaying, which requires a dedicated  time slot for relay transmission. It is worth pointing out that the concept of full duplexing can be applied to non-cooperative NOMA scenarios and is shown to be effective  to improve the spectral efficiency of the joint design of uplink and downlink \cite{jsacnoma32}. 

\subsection{Employing dedicated relays}
Another form of cooperative NOMA  employs  dedicated relays to assist  the NOMA users, as shown Fig. \ref{Fig:plt_CR1}. One motivation for this form of cooperative NOMA is to spectrally efficiently  reach users close to the cell edge. The resulting  benefits are illustrated with the following example. Assume  that there is a dedicated relay which is used to help two users located close to the cell edge. When cooperative OMA is used, four time slots are required for transmission. In particular, it takes two time slots for the BS to deliver the two users' information to the relay, and  another two time slots for the relay to deliver the messages to the two users. With cooperative NOMA,  only  two time slots are required, one for NOMA broadcasting from the BS to the  dedicated relay and the other one for the NOMA transmission from the relay to the two users. The superior spectral efficiency of cooperative NOMA can be immediately deduced  from the fact that  the required number of time slots is reduced from four to two. The benefits of buffer-aided relaying on NOMA with a dedicated relay is investigated in \cite{7803598}.  Note that the idea of cooperative NOMA employing dedicated relays  is general and has been applied to various scenarios with different numbers of transmitters and receivers \cite{7482799,7222419,7417453,7556297, jsacnoma23}.

The other motivation for using dedicated relays is that there can be a large number  of idle users in wireless networks, e.g., in   networks deployed in sport stadiums or convention centers. These idle users can be used as dedicated relays to help other users and to improve   system  coverage. When many dedicated relays are available, a relevant  problem is relay selection, which was studied for the first time for NOMA in  \cite{7482785}. An important conclusion drawn in this work is that the max-min criterion, which is optimal for conventional cooperative networks, is not  optimal for cooperative NOMA. Thus, a two-stage relay selection protocol was proposed, where users are ordered according to their QoS requirements, instead of their channel quality. Particularly, in the first stage of relay selection, relays which can guarantee the performance of the user with strict QoS requirements are identified and grouped into a subset. The second stage is used to select from the qualified relay subset the relay  that yields the largest rate for the other user which only needs to be served opportunistically.  It is proved in \cite{7482785} that this relay selection strategy does not only outperform the max-min scheme, but also  minimizes the overall outage probability.

\section{Millimeter-Wave (mmWave) NOMA}\label{section mmwave noma}
MmWave transmission has   been identified as one of the key enabling technologies for 5G  \cite{6515173}. Both mmWave communications and NOMA are motivated by the   fact that the  spectrum resources below $6$ GHz  available for wireless communications are   limited. Unlike NOMA,  which increases the spectrum efficiency, mmWave utilizes  the less-occupied mmWave frequency bands. In July 2016, the US Federal Communications Commission (FCC) approved that  more than 10 GHz of spectrum in the mmWave bands above $24$ GHz are made  available for 5G wireless communications, a key milestone for mmWave communications \cite{fccmmwave}.

Even though there is a huge amount of spectrum resources available in the mmWave bands, the use of NOMA is still important in mmWave networks for the following two reasons.  Firstly, the application of NOMA in mmWave networks provides an important tool to support massive connectivity. For example, assume  that a mmWave network is deployed in a sports center with thousands of users. The use of NOMA ensures that a huge number  of users with different QoS requirements can be  served simultaneously, which is not possible with OMA. Secondly, the rapid growth of the demand for emerging data services, such as virtual reality and augmented reality, will quickly dwarf the gain obtained from using the mmWave bands. For example, \cite{NUSSBAUM}   predicts that a data rate of $1000$ Gbits/s is required to deliver a high quality telepresence. The use of NOMA can effectively improve the spectral efficiency of mmWave communications, and cope with the rapidly growing demands \cite{jsacnoma13}.

\begin{figure}[thb]
\begin{center}
\includegraphics[width=7cm]{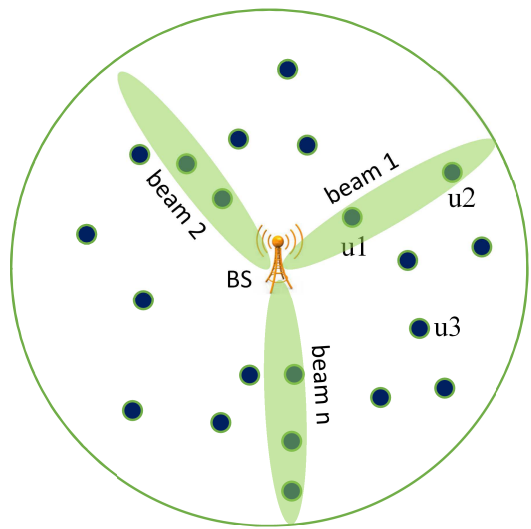}
\end{center}
\caption{Random beamforming in a mmWave-NOMA communication network.  Users, whose channels are strongly correlated, are paired together to share the same beam and perform NOMA.     }
        \label{Fig:plt_CR3}
\end{figure}

It is worth noting that some features of mmWave propagation also facilitate an effective combination of the two 5G technologies. Particularly, the high directionality     of mmWave transmission means that users in mmWave networks may have strongly correlated channels. For example, consider a mmWave network in which the BS is equipped with $M$ antennas and each user has a single antenna. As shown in \cite{7434598} and \cite{7279196}, the $k$-th user's $M\times 1$ channel vector, denoted by $\mathbf{h}_k$,  can be expressed as follows:
 \begin{align}\label{channel model}
\mathbf{h}_k =  \sqrt{M} \frac{a_{k}\mathbf{a}(\theta_{k})}{\sqrt{ 1+d_k^{\alpha}}},
\end{align}
where $M$ is the number of antennas at the BS, $
\mathbf{a}(\theta) = \frac{1}{\sqrt{M}}\begin{bmatrix} 1 & e^{-j\pi \theta} &\cdots &e^{-j\pi(M-1) \theta} \end{bmatrix}^T$, $\theta_k$ denotes the normalized direction  of the line-of-sight (LOS) path,
$d_k$ is the distance between the $k$-th user and the BS,    $\alpha$ is the path loss exponents for the LOS path, and
$a_{k}$ denotes the complex gain for the LOS path. For simplicity of  illustration, we   ignore the non-line-of-sight (NLOS) paths, since their magnitude   can be $20$dB weaker than that of the LOS path   \cite{7279196, 6363891}. In a network with densely deployed users, it can be expected that several  users, such as users $1$ and $2$ shown in Fig. \ref{Fig:plt_CR3}, will share the same normalized LOS direction, which means that their  channels are strongly correlated. In OMA based networks, channel correlation   reduces the multiplexing gain and  the system throughput. However, as illustrated with   the principle of quasi-degradation \cite{7555306}, when users' channel vectors are strongly correlated, the use of NOMA can yield the DPC performance, i.e., the optimal performance for MIMO transmission. In other words,  NOMA is well aligned with the characteristics of  mmWave transmission and leads to substantially improved  system throughput.

In practice,  mmWave transmission is commonly combined with massive MIMO, since the use of mmWave communications makes it possible to pack a large number of antenna elements in a small area. However, the performance gain of  this combination can be realized only if the massive MIMO BS has perfect channel state information (CSI), an assumption that is difficult to realize in practice \cite{6736761}. In \cite{Zhiguo_mmwave}, the use of random beamforming is proposed for mmWave-NOMA networks, in order to reduce the system overhead.  Particularly, random beamforming vectors are generated by the BS, and each user needs to feed back only its scalar  effective channel gain  to the BS, instead of the whole channel vector. Note that the system overhead can be further reduced by exploiting  the directionality  of mmWave transmission. For example, consider a randomly generated beam $\mathbf{p}=\mathbf{a}(\bar{\theta})$, where $\bar{\theta}$ is randomly generated. The effective channel gain of a user for this randomly generated beam, $|\mathbf{h}_{j}^H   \mathbf{p}|^2$, can be written as follows:
\begin{align}
|\mathbf{h}_{j}^H   \mathbf{p}|^2 &=\frac{|a_{j}|^2\left| \sum^{M-1}_{l=0}e^{-j\pi l (\bar{\theta}-\theta_j)}  \right|^2}{{M(1+ d_j^{\alpha})}}\\ \nonumber &=   \frac{|a_{j}|^2   F_M\left(\pi[\bar{\theta} - \theta_j]\right)}{{(1+ d_j^{\alpha})  }},
\end{align}
where $F_M(x)$ denotes the Fej\'er kernel. As shown in \cite{7279196} and \cite{Zhiguo_mmwave}, the behavior of the Fej\'er kernel is very similar to that of the sinc function, i.e., the magnitude of the Fej\'er kernel $F_M\left(\pi[\bar{\theta} - \theta_j]\right)$ is rapidly decreasing for increasing $[\bar{\theta} - \theta_j]$. By exploiting  this phenomenon, it is shown in \cite{Zhiguo_mmwave} that for a given beam not all the users need to feed back their channel conditions. For example,  in Fig. \ref{Fig:plt_CR3}, for beam $1$, only users $1$ and $2$ need to feed back their channel conditions, and there is no need for the other users to feed back their channel gains since their effective channel gains for this beam will be very small.

\section{Practical Implementation  of NOMA}\label{section implementation}
In this section, we discuss a number of important implementation challenges  which have to be addressed before NOMA can be successfully applied in practical wireless systems.
\subsection{Coding and Modulation for NOMA}
Effective  channel coding and modulation schemes are crucial for NOMA, in order  to ensure that the   achievable rates predicted by theory  can be realized in practice. For example, in \cite{7497580} and \cite{7416630}, pulse amplitude modulation (PAM) combined with gray labeling and turbo codes is  applied to NOMA. The resulting new NOMA scheme, which does not rely on SIC,  is shown to be superior to conventional  OMA and NOMA schemes. In addition to turbo codes, other types of channel codes are also   applied to NOMA,    see e.g.\cite{lpma} and \cite{7458388}. The impact of finite-alphabet inputs  on NOMA assisted Z-channels is studied in \cite{jsacnoma30}.

\begin{figure}[thb]
\begin{center}
\includegraphics[width=9cm]{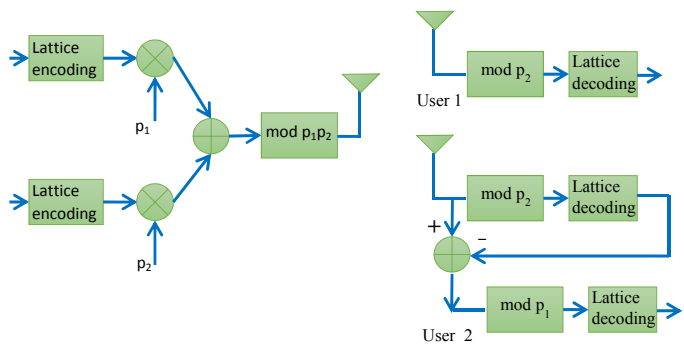}
\end{center}
\caption{An example for LPMA downlink transmission with two users. }
        \label{Fig:plt_CR5}
\end{figure}

More importantly, the integration of sophisticated   coding and modulation with NOMA has also led to the  development of new forms of NOMA, such as Network-Coded Multiple Access (NCMA) \cite{jsacnoma9} and LPMA \cite{lpma}. For   illustration purpose, take   LPMA  as an example. LPMA is based on the property of lattice codes that an integer linear combination of lattice codes is still a lattice code. For a downlink scenario with two users, as shown in Fig. \ref{Fig:plt_CR5}, LPMA     encodes two users' messages  by using lattice coding, such that  the transmitted signal is a linear combination of the two encoded messages which are multiplied with a prime number, respectively,  i.e., the weak user's message is multiplied by a larger prime number, denoted by $p_1$, and the strong user's message is multiplied by a smaller one, denoted by $p_2$, $p_1>p_2$. Multiple access interference is removed by using the modulo operation at the receivers as shown in Fig. \ref{Fig:plt_CR5}, where the weak user employs  a modulo operator with respect to $p_2$ in order to remove the strong user's message. We note that the manner  in which   LPMA   removes multiple access interference is very similar to direct-sequence code division multiple access (DS-CDMA). However, LPMA avoids a severe  disadvantage of CDMA, namely  that the chip rate is much larger than the data rate.  As shown in  \cite{lpma},  LPMA can outperform   conventional power domain NOMA, particularly when the users' channel conditions are similar.

\subsection{Imperfect CSI}
Imperfect CSI  is one of the key obstacles in realizing   the performance gain of NOMA in practice. In general,   different categories of  imperfect CSI are distinguished,  namely  channel estimation errors, partial CSI, and limited channel feedback.
 \begin{itemize}
 \item Channel estimation errors are caused by the imperfect design of channel estimation algorithms and noisy observations, and are damaging to NOMA networks, since these errors result in user ordering ambiguities. The impact of channel estimation errors on power-domain NOMA and SCMA is  studied in \cite{7024794, 7561515 , jsacnoma33}, and the design of pilot transmission for NOMA schemes is investigated in \cite{jsacnoma20}. These   studies have demonstrated that NOMA is actually more resilient to channel estimation errors, compared to OMA. The resource allocation problem is studied in \cite{7561515y} for multi-carrier NOMA when only statistical CSI is available at the transmitter. The proposed design exploits the heterogeneity of the QoS requirements and the statistical CSI to determine the SIC decoding order. The proposed  suboptimal NOMA power allocation and user scheduling scheme achieves a close-to-optimal performance and significantly outperforms OMA.

      \item The use of partial CSI is motivated by the fact that small scale multi-path fading varies much faster than large scale path loss. This means that learning  the path loss information of the users' channels requires   less overhead at the transmitter than estimating  both multi-path fading and path loss. In \cite{7361990,jsacnoma12}, the use of such partial CSI  in NOMA networks is considered, and it is shown that this partial CSI is sufficient for realizing the performance gains that NOMA offers over   OMA.

\item The benefit  of limited feedback is to reduce the system overhead, compared to the case that each receiver feeds all  channel information back to the transmitter \cite{7506136,jsacnoma7}. Let's consider  the one-bit feedback NOMA scheme, proposed in  \cite{7506136}, as an example. The BS broadcasts a threshold to all the users. Each user   compares its received signal strength to this threshold and feeds $1$ (or $0$) back to the BS, if its signal strength is above (or below) the threshold. This one-bit feedback is particularly important for  NOMA networks, when the nodes are equipped with multiple antennas, e.g.,  \cite{Zhiguo_massive} and \cite{Zhiguo_mmwave}. Instead of feeding back the whole vector/matrix, each user only needs to feed back one bit to the BS. Obviously the choice of the threshold is crucial for   one-bit feedback schemes, and   optimal  thresholds are developed in \cite{7506136} for systems  with different power constraints.

\end{itemize}

\subsection{Cross-layer resource allocation}
A typical NOMA network is a complex system, due to the multiple users that have to be served, while the degrees of freedom for  resource allocation, such as user clustering/grouping, power allocation, beamforming/precoding design,  subcarrier allocation, etc.,  are coupled. Even if a centralized resource allocation design  can yield the optimal performance, this scheme entails a prohibitive signaling  overhead and complexity \cite{Robertnoma} and \cite{7587811}. As a result,   distributed  resource allocation in NOMA networks has attracted significant  attention.

In \cite{7523951}, the energy efficiency of NOMA is investigated, where power allocation and subchannel assignments are jointly designed by applying the difference of convex (DC) programming method. In \cite{jsacnoma28}, a queuing  theoretic approach for joint rate control and power allocation is proposed.  In \cite{7557079}, the system throughput maximization problem is studied for both uplink and downlink NOMA, where power allocation and user clustering are  optimized  in an alternating  manner. A similar alternating  approach is also  proposed in \cite{7764327} by employing proportional fairness to decouple the joint design of user scheduling and power allocation.  If only two users are served at each subchannel, the problem of user pairing in the context of NOMA can be  modeled by applying matching theory. For example, each user has a preference list which specifies  with whom this user wants to be paired, and  matching theory can  effectively model  the dynamic interactions among the users. Motivated by this observation, various matching theory based schemes for user pairing have been developed  for NOMA communication scenarios \cite{7560605, lingyangnoma, lingyangnoma2}.

\section{Future Research Challenges }\label{section future}
In this section, a number of promising   directions for future NOMA research are outlined.

 \subsection{The application of wireless power transfer to NOMA}
The motivation for the application of simultaneous wireless information and power transfer (SWIPT), a new member of the energy harvesting family \cite{6489506, 6623062, Dingpoor133}, to NOMA can be illustrated  with the cooperative NOMA scenario considered in \cite{yuanweijsac} as an example. As discussed before, cooperative NOMA can effectively help the user with weak channel conditions, by employing the strong user as a relay. However, in practice, this  user may not want to perform  relaying, since this will consume its own energy and hence shorten its battery life. With SWIPT, the strong user can  harvest energy from the signals sent by the BS, and exploit the harvested energy to power the relay transmission. As a result, the strong user will have more incentive to perform  relaying and help the weak user. Following the idea of \cite{yuanweijsac}, in \cite{7581209}, the transceiver design for   cooperative SWIPT-NOMA is investigated. The achievable rate region of wireless power transfer assisted NOMA is characterized in \cite{jsacnoma31}.   The impact of user selection and antenna selection on cooperative SWIPT-NOMA is studied in  \cite{7752882} and  \cite{7782123}, respectively.

Note that SWIPT is not only applicable to cooperative NOMA, but is also useful for other NOMA communication scenarios. For example, in \cite{7582543, 7510866, 7536861}, SWIPT is applied for NOMA uplink transmission, where users harvest energy from the BS and then send their information to the BS simultaneously by using the NOMA principle. Resource allocation for this form of uplink SWIPT-NOMA transmission is studied in \cite{7585057}, where power allocation and the durations for power and information transfer are jointly designed in order to combat the doubly near-far effect. It is noted that most existing SWIPT-NOMA schemes   rely on  various  idealizing  assumptions, and the impact of practical constraints, such as  hardware impairments, the nonlinear energy harvesting characteristic,  circuit energy consumption, etc., on the performance of SWIPT-NOMA has not been investigated yet.

\subsection{The combination of NOMA and cognitive radio networks}
As discussed in Section \ref{section single carrier}, the application of the cognitive radio concept can significantly reduce the complexity of the design of power allocation policies and strictly guarantee the users'  QoS requirements \cite{Zhiguo_CRconoma, 7527682, 7794658}. The interplay between the two communication concepts is bi-directional, and the application of NOMA is also important to cognitive radio networks. For example, in \cite{7398134}, the NOMA principle is applied to large scale underlay cognitive   radio networks, in order to improve the connectivity of secondary networks. Unlike for  applications of NOMA in conventional  wireless  networks, the power of the   superimposed signals of cognitive radio NOMA users needs to be constrained in order to avoid  excessive  interference to the primary receivers.  In \cite{7740931}, NOMA is employed by the secondary transmitter, which supports two functionalities. One is to deliver information to its own receivers, i.e., the secondary receivers, and the other one is to act as a relay helping the primary receivers.
The current research results  on  the combination of NOMA and cognitive radio networks are still very much dependent  on the considered network topologies, and more work  is needed to gain a fundamental and general understanding of  the synergy  between these two advanced communication techniques.

\subsection{Security provisioning for NOMA}
Similar to other multiple access techniques,   security provisioning  was not  considered when the NOMA principle was developed. A particular security risk with NOMA is that for  SIC to be performed, one user has to be able to  decode the other user's message. It is worth pointing out that such a security risk also exists for other multiple access techniques, e.g., a TDMA user can switch on during a time slot not allocated to it and attempt to decode another user's information. In todays'  telecommunication systems, security is provided by   encryption techniques, instead of relying on multiple access strategies.

However, initial studies have shown that the use of NOMA is helpful in  improving    transmission security, particularly in the following two types of scenarios:
\begin{itemize}
\item For scenarios with external eavesdroppers, the NOMA principle can be combined with physical layer security (PLS) \cite{7426798, 7510755, 7812773, jsacnoma1}. Particularly, the benefit   of NOMA is that the NOMA power allocation coefficients are designed according to the legitimate users' channel conditions, which means that SIC at the eavesdroppers might not be possible and hence eavesdropping may be effectively suppressed.

\item For scenarios in which some NOMA users are potential eavesdroppers, NOMA is shown to be helpful in avoiding  eavesdropping  in \cite{Zhiguo_pls}. Particularly, the NOMA principle is  used in \cite{Zhiguo_pls} to send multicast and unicast messages simultaneously. Beamforming at the base station is carefully designed to artificially enlarge the difference between the two types of users' channel conditions, which is helpful to increase the secrecy data rate.
\end{itemize}

The combination of PLS and NOMA is a rich and promising research area, and more research is needed to develop practical and low complexity schemes for realizing  security in NOMA.

\subsection{Applications of NOMA to other 5G scenarios}
As a promising 5G technique, NOMA has been shown to be compatible with other key enabling techniques for 5G communications. For example, the heterogeneous network architecture will play an important role in 5G networks, where macro base stations and small cell base stations cooperate for spectrum sharing. The benefits of NOMA for heterogeneous networks has been demonstrated  in \cite{jsacnoma3, jsacnoma18,jsacnoma19,jsacnoma25}, as more users can be served in a small cell by exploiting  the NOMA principle. In \cite{jsacnoma4, jsacnoma8,jsacnoma26,jsacnoma29}, the applications  of NOMA to machine-to-machine (M2M) communications, ultra-dense networks (UDN), and massive
machine type communications (mMTC)   are studied, respectively, where the use of NOMA can effectively support massive connectivity and the IoT functionality of 5G.  Content caching is another important technique which has been recognized as a spectrally efficient  way to deliver content to users, and the application of NOMA to content caching is considered in \cite{jsacnoma22}. 

\subsection{Emerging applications of NOMA beyond 5G}
The NOMA principle is quite general and its application is not limited to cellular networks. As discussed in the introduction, the NOMA principle has already been used for TV broadcasting and has been included into the next generation TV standard \cite{7378924}. In addition, the concept of NOMA has attracted significant  attention from the visible light communications (VLC) research community \cite{7275086, 7752879, 7343509, 7572968,7342274}. As outlined in \cite{7275086}, the application of NOMA in VLC systems is beneficial  to support more users, and the fact that VLC offers high SNRs is beneficial  for the application of NOMA, since the performance gap between NOMA and OMA is large in the high SNR regime.  While the existing studies have shown that NOMA-VLC outperforms OMA-VLC, there has been little work on MIMO-VLC. Note that   VLC channels are quite different from conventional radio frequency fading channels, and the users' channels in VLC can be strongly correlated. As indicated by the quasi-degradation criterion in \cite{7555306}, the use of NOMA in scenarios with strongly correlated channels can offer a performance close to the optimal DPC performance. Therefore, a promising future direction in NOMA-VLC is to apply the NOMA principle for the design of VLC precoding/beamforming. Other important applications of NOMA beyond 5G include integrated terrestrial-satellite networks \cite{jsacnoma5}, ALOHA based random access networks  \cite{jsacnoma10}, and vehicular ad-hoc networks \cite{jsacnoma11,jsacnoma17, jsacnoma14}.

\section{Conclusions}\label{section conclusion}
NOMA is an important enabling technology for achieving  the 5G key performance requirements, including high system throughput, low latency, and massive connectivity. As shown in this survey, by exploiting the users' heterogeneous  channel conditions and QoS requirements, NOMA can utilize the scarce bandwidth resources more efficiently than OMA, and existing studies have already clearly demonstrated the ability of NOMA to improve the system throughput. Since multiple users can be served simultaneously, massive connectivity can be realistically achieved with  NOMA, and   NOMA networks  also reduce the delay since users are no longer forced  to wait until an  orthogonal resource block becomes available. The recent industrial efforts to include NOMA in 5G, LTE-A, and digital TV standards demonstrate that NOMA will be an integrated part of  future  generation  wireless networks, and we hope that this survey and the papers in this special issue will be useful to the readers to gain a better understanding of  the benefits and opportunities that  NOMA offers as well as its  practical application scenarios.

\bibliographystyle{IEEEtran}
\bibliography{IEEEfull,trasfer}

  \end{document}